\documentclass[doublecol]{epl2} 
\usepackage{amsmath}
\usepackage{amssymb}
\title{Fixation and Polarization in a Three-Species Opinion Dynamics Model}
\shorttitle{} 

\author{Mauro Mobilia}
\shortauthor{M. Mobilia}

\institute{Department of Applied Mathematics, School of Mathematics, University of Leeds, Leeds LS2 9JT, United Kingdom}


\abstract
{Motivated by the  dynamics of cultural  change and diversity, we  generalize the three-species constrained 
voter model on a complete graph introduced in [J.~Phys.~A {\bf 37}, 8479 (2004)].
In this  opinion dynamics model, a population of size $N$ 
is composed of  ``leftists'' and ``rightists'' that interact with 
``centrists'': a leftist and centrist can both become leftists with rate $(1+q)/2$ or centrists with  rate 
$(1-q)/2$ (and similarly for rightists and centrists), where $q$ denotes the bias towards extremism ($q>0$) or centrism
($q<0$).  This system admits three  absorbing fixed points and
 a ``polarization'' line along which a frozen mixture of leftists and rightists coexist.
In the realm of Fokker-Planck equation, and using a mapping onto a population genetics model,
we compute the fixation probability of ending in every absorbing state
and  the mean times for these events. We therefore
show, especially in the limit of weak bias and large population size
when $|q|\sim N^{-1}$ and $N\gg 1$, how fluctuations alter the mean field predictions: polarization is  likely when $q>0$, but there is always a finite probability to reach a consensus; 
the opposite happens when $q<0$. Our findings are 
corroborated by stochastic simulations.}

\begin{document}

\maketitle

\section{Introduction}
Understanding how diversity is maintained and how traits copied by imitation spread are central issues in genetics, ecology, 
and in behavioral science~\cite{PopGen,Ecology,EGT1,EGT2,Axelrod,BoundedCompromise,SocRev}. 
In this context, there has recently been an upsurge of interest in statistical physics models predicting biological
 and cultural change, see e.g.~\cite{SocRev,voter-variants}, with relevant
phenomena  described by closely related models~\cite{mappings}.

One of the basic issues in opinion dynamics is to understand
the conditions under which consensus or diversity
is reached from an initial population of individuals
(agents) with different opinions. The voter model~\cite{Liggett} is arguably the simplest and most popular 
opinion dynamics model~\cite{SocRev}. In the voter model and in its variants~\cite{voter-variants}, the evolution is  implemented by allowing 
each agent, viewed as a ``spin''~\cite{Liggett}, to adopt a
new state in response to opinions in a local neighborhood. 
While the classic 2-state voter model unavoidably evolves towards consensus,
it has recently been proposed that the competing features of {\it consensus} and {\it incompatibility}
 could be two realistic ingredients to help explain cultural diversity
as an alternative to consensus~\cite{Axelrod,BoundedCompromise}.
The basic idea is that  agents with sufficiently disparate opinions do not interact breaking  up into distinct cultural states 
 and in this case no consensus can be reached (``incompatibility''), while individuals sharing close opinions 
 may evolve towards a global consensus. Influential examples of models characterized by consensus and incompatibility
 are the  Axelrod~\cite{Axelrod} and the bounded
compromise models~\cite{BoundedCompromise} that describe the formation
and evolution of cultural domains.
Recently, a (symmetric) discrete three-state version
of the bounded compromise model was studied in low dimensions (where it exhibits  slow 
non-universal kinetics)~\cite{VKR03}
and solved analytically in its zero-dimensional formulation~\cite{Redner04}.
In such a three-state opinion formation model, there are two species, $A$ and $B$, respectively called 
``leftists'' and ``rightists'', that do not interact among them (incompatibility)~\cite{VKR03,Redner04}. However,
$A$ and $B$ individuals interact with the third species, $C$ (``centrists''), and thus indirectly compete to impose a 
consensus. Due to the $A$ and $B$ incompatibility,  the final state can be either consensus or {\it polarization} with 
a frozen composition of 
leftists and rightists.

It is natural to generalize the three-state constrained voter model of Refs.~\cite{VKR03,Redner04}
by assuming that the interaction between ``extremists'' ($A$ and $B$ individuals) and centrists
is characterized by a bias $q$: extremists are more persuasive when $q>0$, while centrists prevail when
$q<0$.
Here, our goal is to study how fluctuations alter the predictions of the mean field rate equations concerning the system's fate in 
the presence of a small bias $q$.
We thus  determine the ``fixation probability''\cite{FixProb} of each absorbing state of the system 
(comprising $N$ individuals on a complete graph) and compute the average times for these events to occur (mean fixation times).
In fact, while polarization is generally the most probable outcome when $q>0$ and centrism is likely to prevail when $q<0$,
we carefully analyze  the effects of the bias on the system's fate, especially when the population size $N\gg 1$ is large and the bias $|q|\ll 1$ is weak but $Nq={\cal O}(1)$, and determine
 the finite probability that the  final state is a
 consensus when $q>0$ and the frozen stationary state when $q<0$. 
\section{The 3-state constrained voter model}
We consider a population of $N$ individuals on a complete graph, $N_A$ are of species $A$, $N_B$ of type $B$ and $N_C$ of species $C$, with
$N=N_A+N_B+N_C$. The idealized complete graph is used because of its simplicity and analytical tractability.
In the language of the voter model, the species $A$ and $B$ represent ``radical opinions'' 
(e.g. leftists and rightists) while the type $C$ stands for an ``intermediate state'' (e.g. centrists). 
Therefore, $A$ and $B$ individuals interact with $C$ but do not interact among them. Hence,
 species $A$ and $B$ both strive to spread at the expense of $C$.
This system therefore evolves through the interactions that species $C$ has with types $A$ and $B$. 
The latter follow the general prescriptions of the voter model and proceed by imitation: one individual is picked randomly
 and adopts the opinion of one of its random neighbor provided that at least one of the individual is of species $C$.
 In this way, the system's dynamics  at each time increment can be schematically described by the following reactions:
\begin{eqnarray}
  A  C &\to A  A \quad &{\rm with \ rate} \ \frac{1+q}{2} \ ;  \nonumber \\
  A  C &\to C  C \quad &{\rm with \ rate} \ \frac{1-q}{2} \ ; \label{react1} \nonumber \\
  B  C &\to B  B \quad &{\rm with \ rate} \ \frac{1+q}{2} \ ; \nonumber \\
  B  C &\to C  C  \quad &{\rm with \ rate} \ \frac{1-q}{2} \ , \label{react2}
\end{eqnarray}
where $-1\leq q\leq 1$. The special case $q=0$ was thoroughly studied 
in Ref.~\cite{Redner04} and will therefore not been discussed here.
The parameter $q$ measures the bias towards polarization (when $q>0$) or centrism (when $q<0$).
In the former situation  extremists ($A$'s and $B$'s) are more persuasive than centrists ($C$'s), 
while in the latter centrism is the dominating and more persuasive opinion. As a consequence,
we anticipate that polarization is the most probable final state when $q>0$, while centrism consensus
is expected to be the most likely stationary state when $q<0$.

At mean field level, assuming a population of infinite size ($N\to \infty$) and the absence of any random fluctuations, the system dynamics is described by the following rate equations (REs) for the densities $a\equiv N_A/N$ and $b\equiv N_B/N$ of species $A$ and $B$, respectively:
\begin{eqnarray}
 \label{RE}
\frac{d}{dt}a=qa(1-a-b),\;
\frac{d}{dt}b=qb(1-a-b),
\end{eqnarray}
where we have used the fact that $a(t)+b(t)+c(t)=1$. 
The REs (\ref{RE}) admit 3 absorbing fixed points,
$(a,b,c)\equiv (a(\infty),b(\infty),c(\infty))\in$
$\{{\cal A}=(1,0,0), {\cal B}=(0,1,0), {\cal C}=(0,0,1)\}$
 and a {\it line of fixed points} given by
${\cal AB}=(a,1-a,0)$, with $0<a<1$. In fact, the system (\ref{RE}) can be solved exactly, yielding~\cite{Redner04}
\begin{eqnarray*}
 \label{MFsol}
a(t)=\frac{x e^{qt}}{1-(x+y)(1-e^{qt})}, \; 
b(t)=\frac{y e^{qt}}{1-(x+y)(1-e^{qt})}, 
\end{eqnarray*}
where $x,y$ respectively are the initial densities of species $A$ and $B$, i.e. $x=a(0)$ and $y=b(0)$.
These results reveal that the line of fixed points is the stable solution when $q>0$, i.e. in this case 
$(a,b,c)=(x/(x+y),y/(x+y),0)$, with $a+b=1$. On the other hand, when $q<0$ and 
the bias favors the species $C$, one finds $a=b=0$ and  $c=1$.
Furthermore, we notice that the REs (\ref{RE}) predict that the ratio of the densities
 is conserved, i.e. $a(t)/b(t)=x/y$.

When the population size $N$ is finite, demographic fluctuations  can drastically alter 
the mean field predictions. In such a setting, the dynamics is no longer deterministic and 
the system's fate depends non-trivially on the bias strength and on the population initial composition,
parametrized by the initial densities $(x,y)$.
Within  a  stochastic formulation of the model, the  moves (\ref{react1}) define a birth-death 
process~\cite{Gardiner}, where, according to (\ref{react1}), the number of individuals of each 
species increases or decreases by one unit in each time step. 
A quantity that is central for our discussion is $P^{{\cal AB}}(x,y)$, the polarization fixation probability
 along the absorbing 
line ${\cal AB}$ starting from an initial population composition $(x,y,1-x-y)$. 
Hence, $P^{{\cal AB}}$ gives the probability to find the system locked into 
a {\it polarized} state where ``extremists'' ($A$'s and $B$'s) coexist without interacting. This probability obeys the following backward 
master equation~\cite{Gardiner}:
\begin{eqnarray}
 \label{backME}
&&(T_x^+ + T_x^- + T_y^+ + T_y^-)P^{{\cal AB}}(x,y)=T_x^{-}P^{{\cal AB}}(x-\delta,y)\nonumber\\ &+&T_x^{+}P^{{\cal AB}}(x+\delta,y)
+ T_y^{-}P^{{\cal AB}}(x,y-\delta)+T_y^{+}P^{{\cal AB}}(x,y+\delta),\nonumber\\
\end{eqnarray}
where, in the limit $N\gg 1$, the transition rates are $T_x^{\pm}\equiv (1\pm q)x(1-x-y)/2$ and 
$T_y^{\pm}\equiv(1\pm q)y(1-x-y)/2$, with $\delta=N^{-1}$. 
This {\it two-dimensional} equation has to be supplemented by the boundary conditions: $P^{{\cal AB}}(x,0)=P^{{\cal AB}}(0,y)=0$ and $P^{{\cal AB}}(x,1-x)=1$. 
By Taylor-expanding (\ref{backME}) to second-order in $\delta$, one finds 
\begin{eqnarray}
\label{EqPAB}
\left\{s[x\partial_x+y\partial_y]+\frac{1}{2}[x\partial_x^2+y\partial_y^2]\right\}P^{{\cal AB}}(x,y)=0,
\end{eqnarray}
where we have introduced the parameter $s\equiv Nq$. 
When $|s|\gg 1$, the diffusion term on the left-hand-side of (\ref{EqPAB})
is 
negligible in front of the deterministic drift term, while the opposite occurs when  $|s|\ll 1$. 
This implies that the interesting situation arises when $s$ is of order one, 
i.e. when $N \gg 1$ and $|q|\sim N^{-1}$; otherwise one would essentially recover
the mean field predictions of (\ref{RE}) when $|s|\gg 1$, or the results of Ref.~\cite{Redner04} when $|s|\ll 1$.
It is also useful to notice that the associated backward  Fokker-Planck (FP)
differential operator ${\cal L}_{{\rm bFP}}$ reads~\cite{Gardiner}:
\begin{eqnarray}
\label{backFP}
{\cal L}_{{\rm bFP}}&=&\frac{(1-x-y)}{2N}\left[2s(x\partial_x  + y\partial_y)
+x\partial_x^2  + y\partial_y^2\right]
\end{eqnarray}
A {\it one-dimensional} version of ${\cal L}_{{\rm bFP}}$ often appears 
in population genetics~\cite{PopGen} and  evolutionary game theory~\cite{EGT2}.
\section{Fixation Probabilities}
 We have seen that the mean field treatment (\ref{RE}) predicts that the line ${\cal AB}$ and 
the fixed point ${\cal C}$ are the system's attractor when $q>0$ and $q<0$, respectively.
The quantity
$P^{{\cal AB}}$, that obeys~(\ref{EqPAB}), and 
the probability density $F^{{\cal AB}}_a(x,y)$ that the system's final 
state has coordinate $(a,1-a)$ along the absorbing line, 
are central to study the system's fate in the presence of fluctuations.
These quantities are related by 
$P^{{\cal AB}}(x,y)=\int_{0}^{1} da F^{{\cal AB}}_a(x,y)$. Hence,  
$F^{{\cal AB}}_a(x,y)$  obeys the same backward FP equation (\ref{EqPAB}) as $P^{{\cal AB}}$ but with 
the boundary conditions $F^{{\cal AB}}_a(0,y)=F^{{\cal AB}}_a(x,0)=0$ and $F^{{\cal AB}}_a(x,1-x)=\delta(a-x)$.
As in Ref.~\cite{Redner04}, Eq.~(\ref{EqPAB}) turns out to be {\it separable} and  {\it exactly solvable}.
To obtain its solution, it is useful to introduce the polar coordinates $(\rho,\theta)$ such that 
$\sqrt{x}=\rho\cos{\theta}$ and $\sqrt{y}=\rho\sin{\theta}$. In these coordinates, 
with $0\leq \rho\leq 1$ and $0\leq \theta\leq \pi/4$,
 Eq.~(\ref{EqPAB})
becomes: 
\begin{eqnarray}
\label{PDE_F}
&&\left[(4s\rho-
\rho^{-1})
\partial_{\rho}+\partial_{\rho}^2
\right]P^{{\cal AB}}(\rho,\theta)
\\
&+&\rho^{-2}
\left(\{\tan{\theta}-\cot{\theta}\}\partial_{\theta}+
\partial_{\theta}^2
\right)
P^{{\cal AB}}(\rho,\theta)=0.\nonumber
\end{eqnarray}
with boundary conditions $P^{{\cal AB}}(\rho=0,\theta)=0$ and $P^{{\cal AB}}(\rho=1,\theta)=1$. 
The probability density $F^{{\cal AB}}_a$ also obeys the equation (\ref{PDE_F}) but with the
boundary conditions $F^{{\cal AB}}_a(\rho=0,\theta)=0$ and $F^{{\cal AB}}_a(\rho=1,\theta)=\delta(a -\cos^2{\theta})$.
Following Ref.~\cite{Redner04}, we
 write $P^{{\cal AB}}$
and
$F^{{\cal AB}}_a$ in the form $\sum_{n} c_n R_n(\rho)u_n(\theta)\sin{(2\theta)}$, where the 
$R_n(\rho)$ and $u_n(\theta)$ are the eigenvectors 
(with eigenvalues  $\lambda_n$)
of the following radial and angular Sturm-Liouville problems, respectively: 
\begin{eqnarray}
\label{SL1}
&&\rho^2 \frac{d^2 R_n}{d\rho^2} + \rho \frac{d R_n}{d\rho} [4s\rho^2 -1]-\lambda_n R_n =0\\
&& \frac{d^2 u_n}{d\theta^2} -\frac{3}{4} \left(\frac{1}{\sin^2{\theta}} + \frac{1}{\cos^2{\theta}}\right)u_n+(1+\lambda_n)u_n =0\nonumber
\end{eqnarray}
\begin{figure}
\begin{center}
\includegraphics[width=3.4in, height=2.0in,clip=]{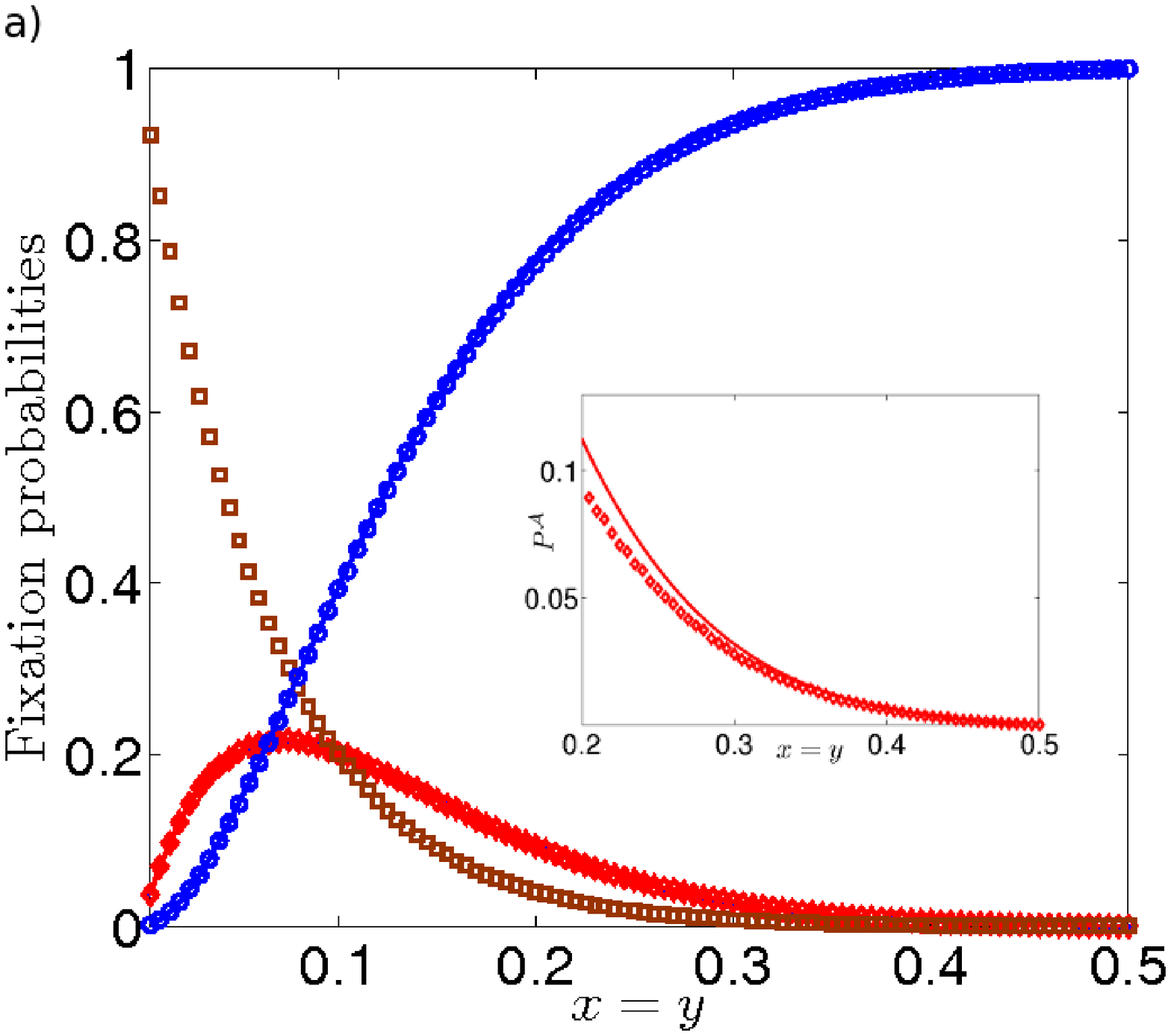}
\includegraphics[width=3.4in, height=2.0in,clip=]{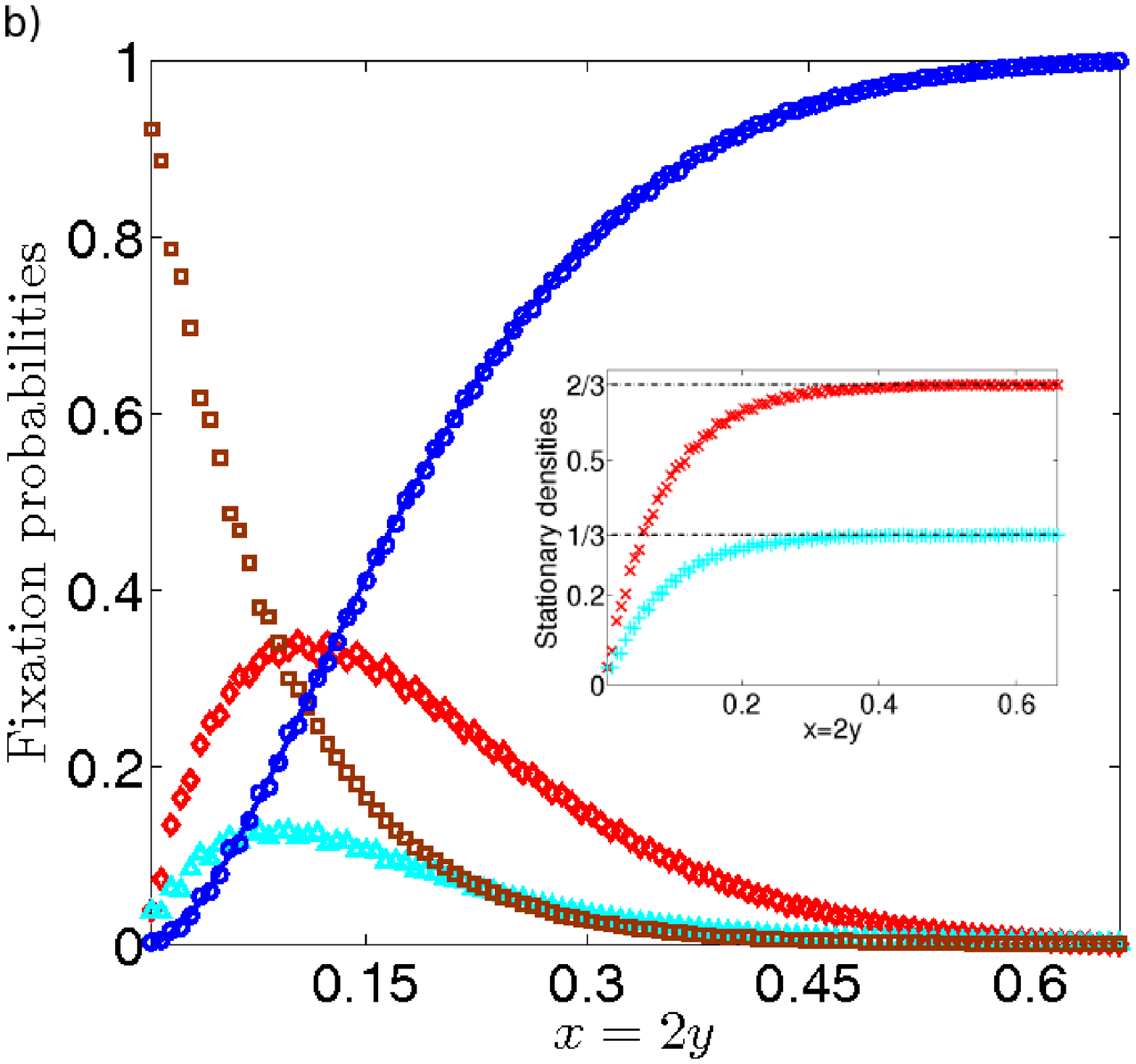}
\caption{{\it (Color online)}.
Fixation probabilities for $s>0$ as functions of $x$: $P^{{\cal A}}$ ($\diamond$), $P^{{\cal B}}$ ($\triangle$), $P^{{\cal C}}$ ($\square$); $P^{{\cal AB}}$ ($\circ$) is compared 
with (\ref{Pab}) (solid curve). 
 Parameters are $N=200, s=4$ (i.e. $q=0.02$). (a) Initial density of $A$ is $x=y$ and $P^{{\cal A}}=P^{{\cal B}}$. Inset: Comparison of $P^{{\cal A}}$
with its analytical approximation, see text. (b) Initial density of $A$ is $x=2y$.
Inset: final densities of species $A$ ($\times$) and $B$ ($+$) as functions of $x=2y$.
 Numerical results have been averaged over  $2\times 10^5$ samples.}
  \label{FixProb_qpos}
\end{center}
\end{figure} 
The
solution to the radial equation (\ref{SL1}) with boundary condition $R_n(0)=0$ reads $R_n(\rho)\propto e^{-s\rho^2} \rho^{-2n-1} I_{n+1/2}(s\rho^2)$, 
where $I_n$ is the modified Bessel function of first kind and order $n$~\cite{Abramowitz}.
The equation (\ref{SL1}) for the angular function $u(\theta)$ coincides with a stationary 
Schr\"odinger equation in a P\"oschl-Teller potential hole  whose solution is $u_n(\theta)\propto \sqrt{\sin{(2\theta)}} P_{n+1}^{1}(\cos{(2\theta)})$,
where $P_{n}^{1}$ denotes an associate Legendre polynomial of first order~\cite{Flugge,Redner04}.
The eigenvalues are found to be 
 $\lambda_n=4(n+1)(n+2)$ and
the coefficients $c_n$ are determined using the orthogonality of the $P_{n}^{1}$'s 
together with the boundary conditions for $F^{{\cal AB}}_a$ and $P^{{\cal AB}}$ ~\cite{Redner04}. 
This leads to
\begin{eqnarray}
\label{Fab}
&& F^{{\cal AB}}_a(x,y)=\sqrt{\frac{xy}{x+y}}e^{s(1-x-y)}\,\sum_{n=1}^{\infty}\left(\frac{2n+1}{n(n+1)}\right)
\nonumber\\
&\times&
\left\{
\frac{I_{n+1/2}(s(x+y))}{I_{n+1/2}(s)}
\right\}
\,P_{n}^{1}\left(\frac{x-y}{x+y}\right)\,
\frac{P_{n}^{1}(2a-1)}{\sqrt{a(1-a)}}. 
\end{eqnarray}
Since $P^{{\cal AB}}(x,y)=\int_{0}^{1}da\,F^{{\cal AB}}_a(x,y)$, using  
the properties of the  associate Legendre polynomials~\cite{Abramowitz}, one also obtains
\begin{eqnarray}
\label{Pab}
P^{{\cal AB}}(x,y)&=&2\sqrt{\frac{xy}{x+y}}e^{s(1-x-y)}\, \sum_{n \,{\rm odd}}^{\infty}\left(\frac{2n+1}{n(n+1)}\right)
\nonumber\\
&\times&
\left\{
\frac{I_{n+1/2}(s(x+y))}{I_{n+1/2}(s)}
\right\}\,P_{n}^{1}\left(\frac{x-y}{x+y}\right),
\end{eqnarray}
where the subscript ``$n \, {\rm odd}$'' means that the sum runs over odd integers $n$. Since $P^{{\cal AB}}$ is the polarization fixation 
probability, the probability that the system's final state is consensus (with either $A, B$, or $C$) is $1-P^{{\cal AB}}$.

The quantities  $P^{{\cal AB}}$ and the fixation probabilities of the  absorbing states ${\cal A}$, ${\cal B}$ and ${\cal C}$, respectively 
denoted $P^{{\cal A}},P^{{\cal B}}$ and $P^{{\cal C}}$, are related by $P^{{\cal A}}(x,y)+P^{{\cal B}}(x,y)+P^{{\cal AB}}(x,y)=a+b=1-P^{{\cal C}}(x,y)$. In fact, as $C$'s interact identically with $A$'s and $B$'s, a mapping onto a population genetics model where extremists are regarded as mutants of a single class   yields $P^{{\cal C}}=\frac{e^{-2s(x+y)}-e^{-2s}}{1-e^{-2s}}$ (see below).
Furthermore,  the fixation probabilities  and 
the species density are related by
by $a=P^{{\cal A}}(x,y)+\int_0^1 da'\,a' F_{a'}(x,y)$
and $b=P^{{\cal B}}(x,y)+P^{{\cal AB}}(x,y)-\int_0^1 da'\,a' F_{a'}(x,y)$,  
with $P^{{\cal B}}(x,y)=P^{{\cal A}}(y,x)$.

When $|s| \ll 1$,
$I_{n+1/2}(s(x+y))/I_{n+1/2}(s)\approx (x+y)^{n+1/2}$~\cite{Abramowitz} and, with (\ref{Fab}) and (\ref{Pab}), 
one obtains 
$\frac{F^{{\cal AB}}_{a;q\neq 0}(x,y)}{F^{{\cal AB}}_{a;q=0}(x,y)}=
\frac{P^{{\cal AB}}_{q\neq 0}(x,y)}{P^{{\cal AB}}_{q=0}(x,y)}= 
[1-s\{x+y-1\}+{\cal O}(s^2)].$
From this expression and the compact result $P^{{\cal AB}}_{q=0}(x,x)=1-\frac{1-4x^2}{\sqrt{1+4x^2}}$ obtained in~\cite{Redner04}, 
we infer 
 $P^{{\cal AB}}(x,x)\approx 
6(1+s)x^2$ when $x=y\ll 1$. Hence, at low initial density $x=y$ and for weak bias ($|s| \ll 1$),
$P^{{\cal AB}}$ is a quadratic polynomial in $x$ 
 with an amplitude  proportional to $s$.
\subsection{Fixation probabilities when $s>0$}
When $s>0$, we find that the fixation probability $P^{{\cal AB}}(x,y)$ displays a sigmoid shape,
that steepens when $s$ increases, interpolating monotonically between 
$0$ (when $x+y\ll 1$) and $1$ (for $x+y\to 1$), see Fig.~\ref{FixProb_qpos}. The agreement between the analytical prediction (\ref{Pab}) and the results
 of stochastic simulations 
(obtained using the Gillespie algorithm~\cite{Gillespie}) is excellent, as illustrated in Fig.~\ref{FixProb_qpos}.
 This figure also shows that $P^{{\cal C}}$ 
steeply decays to zero when $x+y$ increases, as expected from its analytical expression. Furthermore, $P^{{\cal A}}$ and $P^{{\cal B}}$ display positive skewness and 
maxima around the initial densities $(x_*,y_*)$ such that $P^{{\cal AB}}(x_*,y_*)=P^{{\cal C}}(x_*,y_*)$
($x_* + y_*\approx 0.157$ in Fig.~\ref{FixProb_qpos}(a)). 
Moreover, as shown in the inset of 
Fig.~\ref{FixProb_qpos}(b),
the stationary densities steadily approach their mean field values, i.e. $a\to x/(x+y)$ and $b\to x/(x+y)$, when 
$x+y\gtrsim 0.30$. In such a regime, one thus has $a\approx \frac{x}{y+y}=
P^{{\cal A}}(x,y)-\int_0^1 da'\,a' F_{a'}(x,y)$ and, with (\ref{Fab}), this yields the following approximation for 
the fixation probability of $A$:
$P^{{\cal A}}(x,y)\approx \frac{x}{x+y}-\sqrt{\Delta}e^{s(1-x-y)}\sum_{n=1}^{\infty}\left(
\frac{2n+1}{n(n+1)}\left\{
\frac{I_{n+1/2}(s(x+y))}{I_{n+1/2}(s)}
\right\}
\right)P_{n}^{1}\left(\Delta\right)$, where $\Delta\equiv (x-y)/(x+y)$. 
The  inset of Fig.~\ref{FixProb_qpos}(a), shows that the agreement between this approximation and the results of 
numerical simulations increases with $x=y\gtrsim 0.20$. A similar approximation gives $P^{{\cal B}}\approx 1-P^{{\cal AB}}(x,y)-P^{{\cal A}}(x,y)$.
\begin{figure}
\begin{center}
\includegraphics[width=3.045in, height=2.1in,clip=]{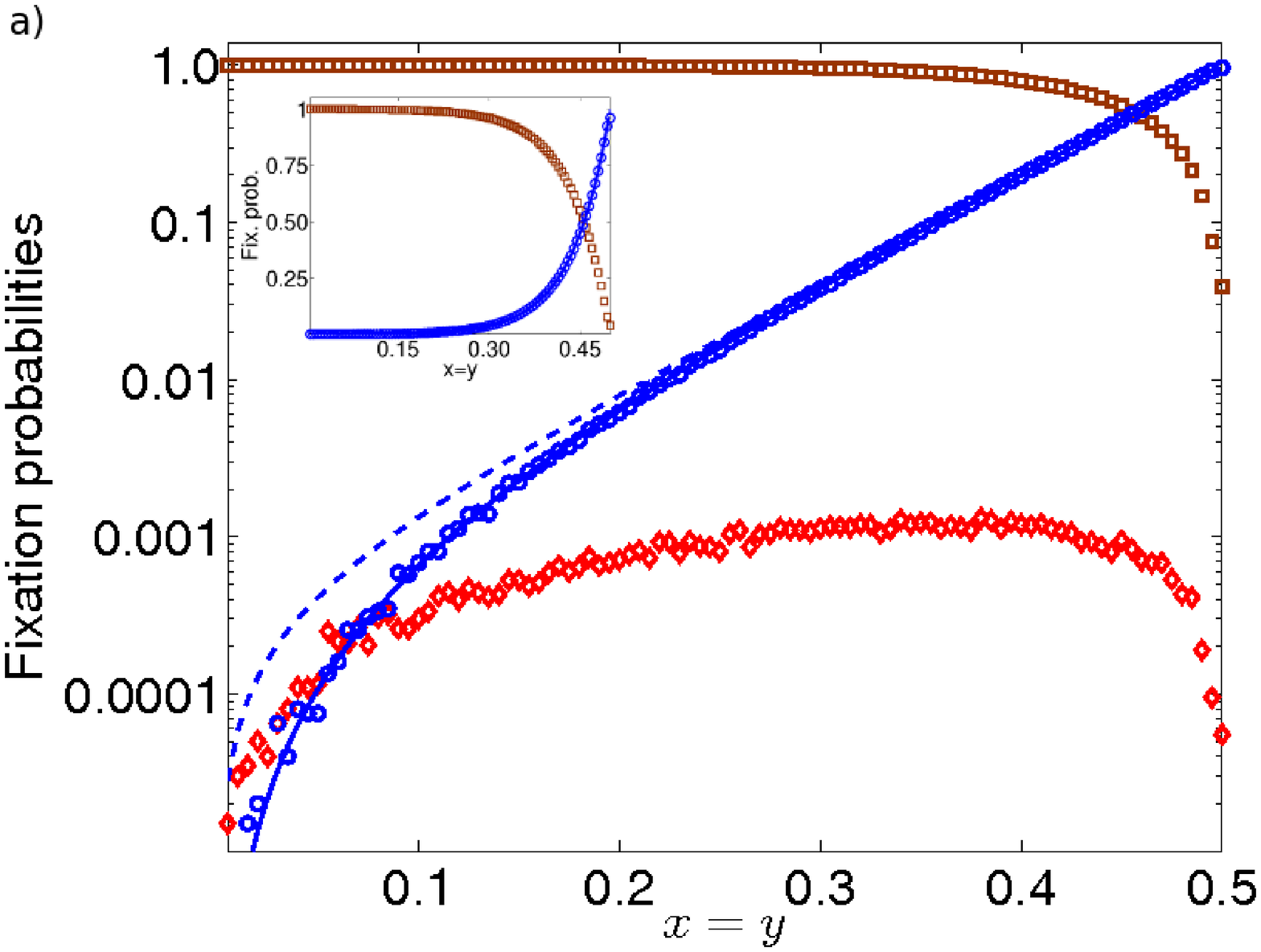}
\includegraphics[width=3.045in, height=2.1in,clip=]{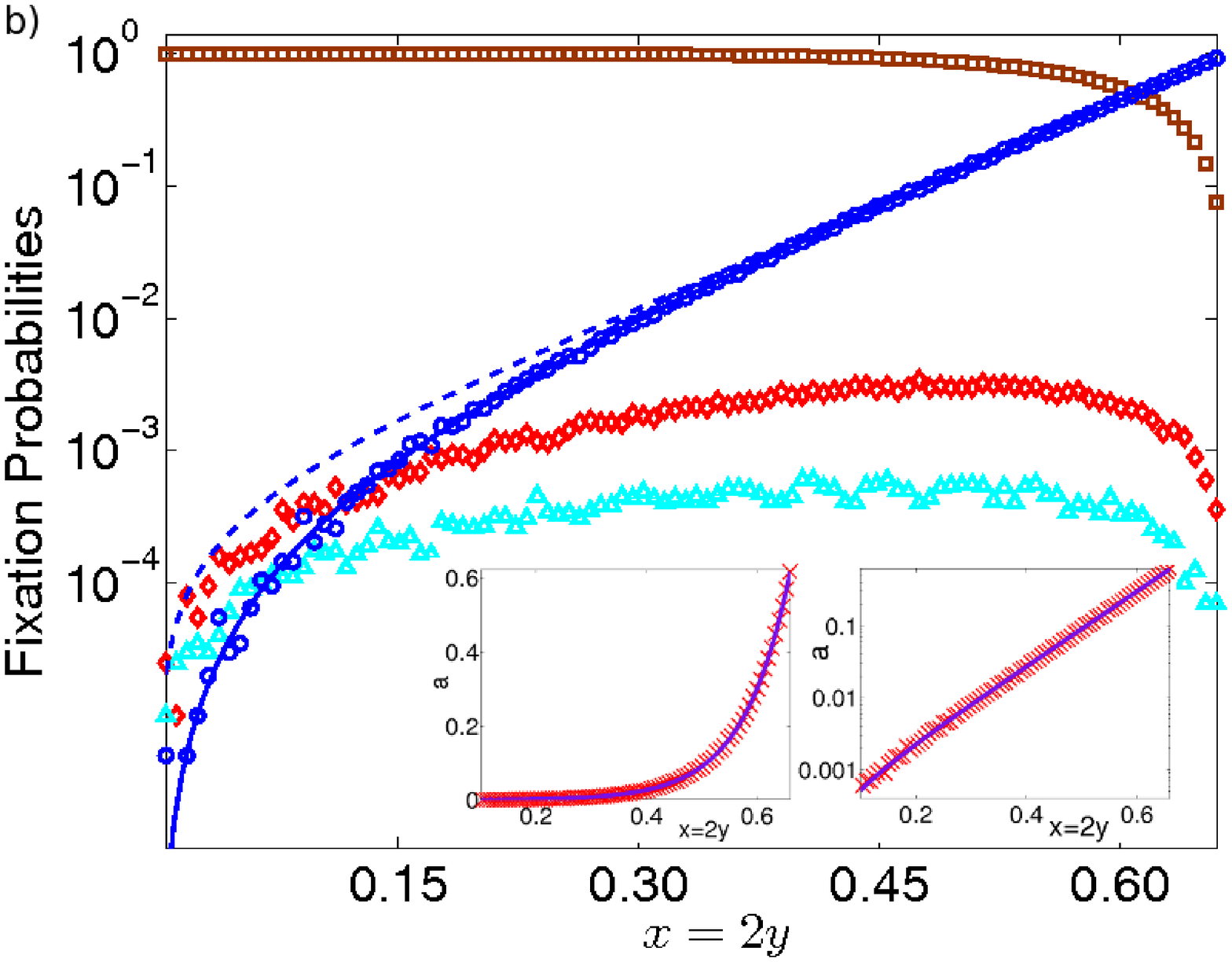}
\caption{{\it (Color online)}. Same as in Fig.~\ref{FixProb_qpos}, with $s=-4$   in semi-log scale, and comparison 
with (\ref{Pab}) (solid curve) and (\ref{Pab-ap2}) (dashed curve), see text.
The initial density of $A$ is $x=y$ in (a) and $x=2y$ in (b). Inset of (a): $P^{{\cal C}}$ and $P^{{\cal AB}}$ 
in linear scale. Insets of (b): stationary density of $A$ ($\times$)  
as function of $x=2y$ compared with the analytical expression (solid curves) given in the text,
in linear (left) and semi-log (right) scales.
}
 \label{FixProb_qneg}
\end{center}
\end{figure}
\subsection{Fixation probabilities when $s<0$}
When $s<0$, the bias is towards the absorbing state ${\cal C}=(0,0,1)$
 and the fixation probabilities  $P^{{\cal A}}$ and $P^{{\cal B}}$ are vanishingly small.
For small initial densities of $A$ and $B$ (i.e. $x+y\ll 1$), the probability of reaching 
the absorbing line ${\cal AB}$ is also very small ($\lesssim N^{-1}$). However, this fixation probability $P^{{\cal AB}}$ grows monotonically when 
$x+y$ increases and, according to (\ref{Pab}) and as shown in Fig.~\ref{FixProb_qneg}(a,b), approaches the value one when $x+y \to 1$.
Figure \ref{FixProb_qneg} shows that  the system's most likely fate is either to end up in 
the state ${\cal C}$  or on the line ${\cal AB}$.
To determine a concise and accurate approximation of $P^{{\cal AB}}$, one can use a mapping onto a 
population genetics model where $C$ is considered as
a ``wild-type'' allele and both
 $A$ and $B$ species are regarded as forming a single-class deleterious allele~\cite{PopGen}. 
In this situation, the probability $P^{{\cal AB}}(x,y)=P^{{\cal AB}}(x+y)$
corresponds to the fixation of the (single-class) ``mutants'' and 
depends only on  $x+y$ and $s$. 
The angular dependence therefore drops out from (\ref{PDE_F})  and, with the 
(new) boundary conditions $P^{{\cal AB}}(x+y=0)=0$ and $P^{{\cal AB}}(x+y=1)=1$, one obtains
the following approximation for $P^{{\cal AB}}$:
\begin{eqnarray}
\label{Pab-ap2}
P^{{\cal AB}}(x+y)\simeq \frac{e^{2|s|(x+y)}-1}{e^{2|s|}-1}.
\end{eqnarray}
As shown in Fig.~\ref{FixProb_qneg},
this result is found to be an excellent approximation of $P^{{\cal AB}}$ when 
$x+y\gtrsim 0.30$. In such a regime  $P^{{\cal A}}\approx P^{{\cal B}} \approx 0$
and $P^{{\cal AB}}(x+y) \simeq 1-P^{{\cal C}}(x+y)$
(see inset of Fig.~\ref{FixProb_qneg}(a)). Using (\ref{Pab-ap2}), one can  obtain 
 the stationary densities $b\simeq a(y/x)$ and
$a \simeq \left(\frac{x}{x+y}\right)\left(\frac{e^{2|s|(x+y)}-1}{e^{2|s|}-1}\right)$. The excellent
 agreement between this  expression and numerical simulations is demonstrated in
the insets of  Fig.~\ref{FixProb_qneg}~(b).
\section{Mean Fixation times} 
\begin{figure}
\begin{center}
\includegraphics[width=3.2in, height=2.2in,clip=]{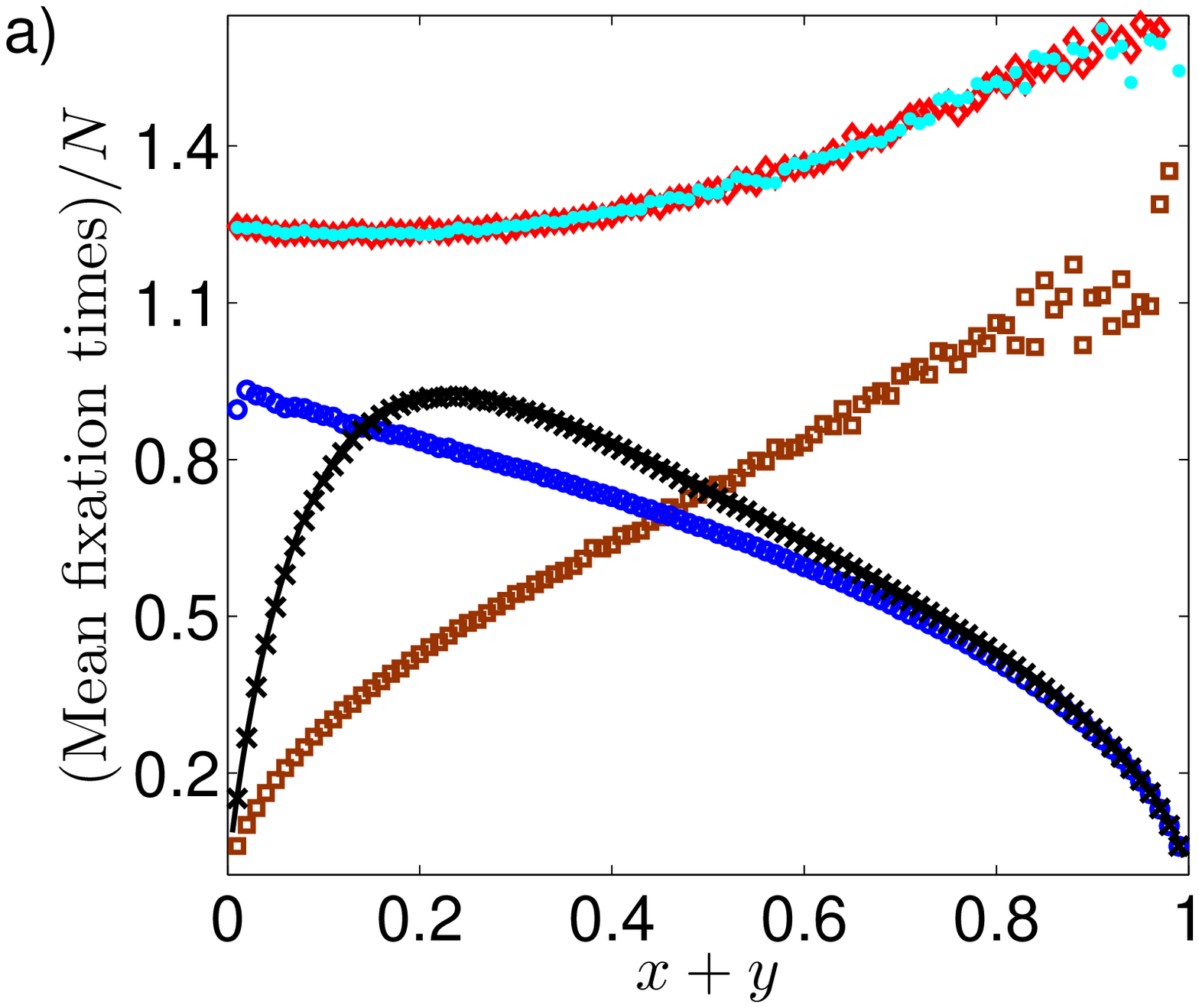}
\includegraphics[width=3.2in, height=2.2in,clip=]{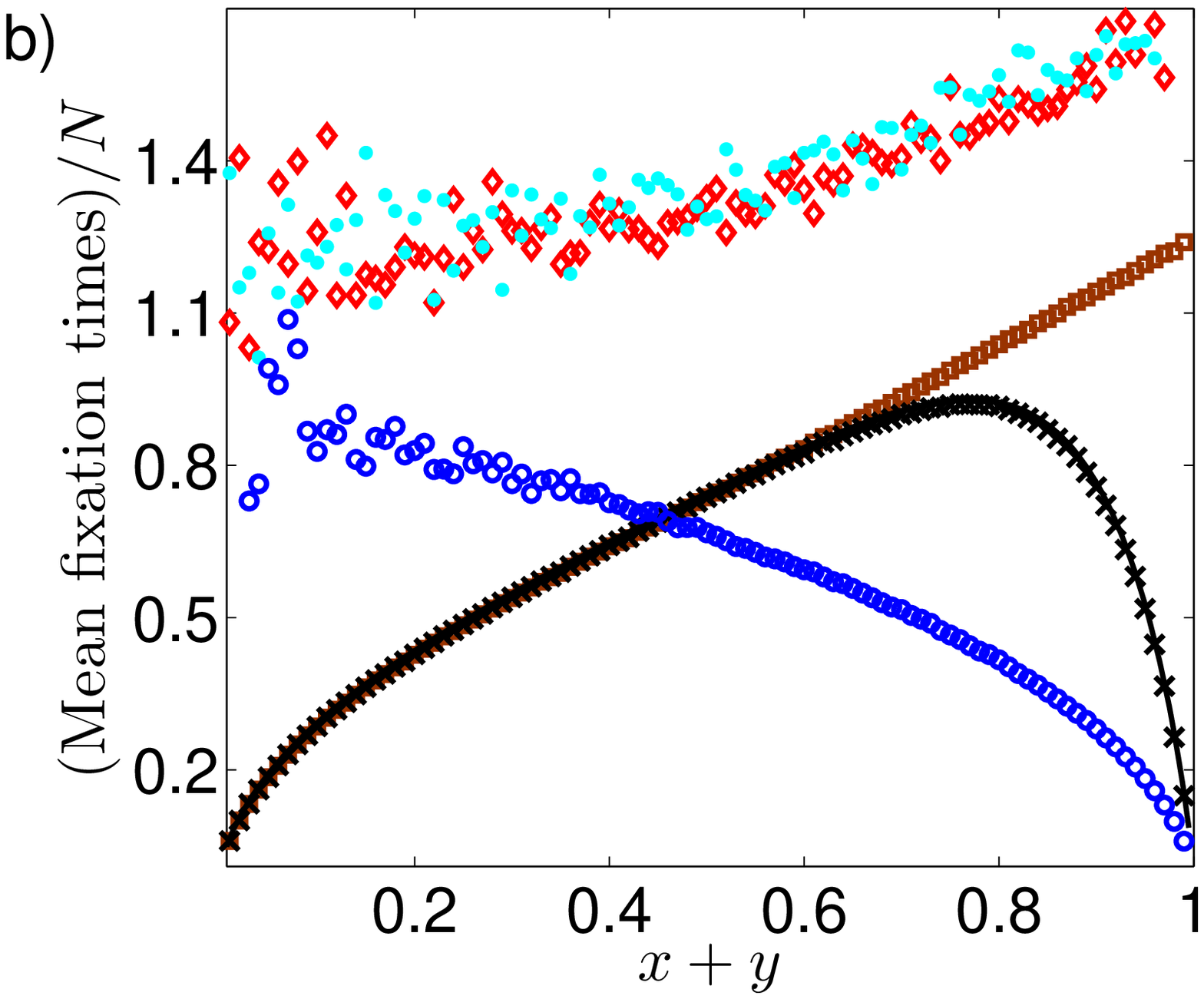}
\caption{{\it (Color online)}. The normalized
unconditional mean fixation time (MFT) $\tau/N$ ($\times$), compared with the solution of Eq.~(\ref{MFTeq}) (solid curve), and the (normalized) conditional MFTs
$\tau^{{\cal A}}/N$ ($\diamond$), $\tau^{{\cal B}}/N$ ($\bullet$), $\tau^{{\cal C}}/N$ ($\square$), $\tau^{{\cal AB}}/N$ ($\circ$), see text. 
The parameters are $N=200$, $x=y$ (initial $A$ density), $s=4$ in (a) and $s=-4$ in (b).  
The maxima of the unconditional MFTs are $\tau(0.22)\approx 0.94N$ in (a) and $\tau(0.78)\approx 0.94N$ in (b).
The numerical results have been averaged over $2\times 10^5$ samples.
}
  \label{FixTime}
\end{center}
\end{figure}

The mean times  necessary to reach the absorbing states, also called the mean fixation times (MFTs),
are quantities of great interest in evolutionary dynamics~\cite{PopGen,EGT1,EGT2,SocRev,voter-variants}. 
Here, the unconditional MFT to reach {\it any} of the system's absorbing states is 
a function  $\tau(x,y)$ of the initial density $(x,y)$ obeying the backward FP equation 
${\cal L}_{\rm bFP}(x,y)\tau(x,y)=-1$ [see (\ref{backFP})] with boundary conditions
$\tau(1,0)=\tau(0,1)=\tau(0,0)=\tau(a,1-a)=0$~\cite{Gardiner,PopGen}.
Such an equation can be solved by 
using an exact mapping onto a suitable population genetics model. In fact, since we are interested in
the {\it unconditional} MFT, and species $C$ does not make any distinction between $A$ and $B$ 
individuals, as seen above, the latter can be considered as belonging  to a single ``mutant type'' class that interacts  with 
the ``wild type'' $C$~\cite{PopGen}. In this formulation, the equation for $\tau$
becomes one-dimensional  with two absorbing states: either only wild-type individuals (consensus with $C$)
or only mutants (either $A$ or $B$ consensus, or polarization in a frozen mixture of $A$'s and $B$'s).
 The unconditional MFT thus depends only on the  initial density $x+y$ of extremists ($A$'s and $B$'s): 
$\tau(x,y)=\tau(x+y)$. Hence, with the variable $w\equiv x+y$, one obtains:
\begin{eqnarray}
\label{MFTeq}
\frac{w(1-w)}{N}\left[2s\frac{d\tau(w)}{dw} + \frac{d^2 \tau(w)}{dw^2} \right]=-1,
\end{eqnarray}
with $\tau(0)=\tau(1)=0$. 
Eq.~(\ref{MFTeq}) frequently appears in population genetics, where it describes the MFT in a diallelic haploid population
in the presence of selection of (weak) intensity $|q|=|s|/N\ll 1$~\cite{PopGen} (see also~\cite{EGT2}).  
An interesting property of (\ref{MFTeq}) is its invariance under the transformation $(w,s)\to (1-w,-s)$. This implies that $\tau(w)$ for $s>0$ coincides with $\tau(1-w)$ for $s<0$, see Fig.~\ref{FixTime}.
 Equation (\ref{MFTeq}) can be solved by standard means (see e.g. \cite{Gardiner,EGT2} and references therein) 
 yielding a cumbersome expression.
The latter takes a more compact form for the special value $w=1/2$, when it reads:
 \begin{eqnarray}
\label{MFT}
&&\tau=\frac{N}{s(1+e^s)}\left[e^{2s}{\rm Ei}(-2s)+(e^s -1)\ln{(2|s|)}\right]
\nonumber\\
&+&\frac{N}{2s e^s(1+e^s)}
\left[e^s (e^s-1)\gamma_{{\rm E}}+{\rm Ei}(s) -{\rm Ei}(2s)
\right]\nonumber\\
&+&
\frac{N}{2s(1+e^s)}\left[2{\rm Ei}(s)  
-2e^s (e^s+1){\rm Ei}(-s) 
\right],
\end{eqnarray}
where $\mbox{Ei}(x)\equiv\int_{-\infty}^x\frac{e^t}t\,dt$ denotes the exponential integral and $\gamma_{{\rm E}}=0.5772...$ 
is Euler-Mascheroni's constant~\cite{Abramowitz}.
From the general expression of $\tau(x+y)$, one finds that the unconditional MFT scales linearly with the 
system size $N$, i.e. $\tau(x+y)=Nf_{\tau}(s,x+y)$, where $f_{\tau}$ is a scaling function. 
Such a scaling relationship is fully confirmed by the numerical results reported in Fig.~\ref{FixTime}(a,b).
We can infer from the latter that $f_{\tau}$ has an inverted u-shape dependence on $w=x+y$
and skewness towards small values of $w$ for $s>0$ and large values of $w$ for $s<0$, see Fig.~\ref{FixTime}(a,b).
This is a consequence of the  $(w,s)\to (1-w,-s)$ symmetry obeyed by $\tau(x+y)$.
 The {\it conditional} mean fixation times, $\tau^{\cal S}$, to reach the absorbing states ${\cal S}\in({\cal A},{\cal B},{\cal C},{\cal AB})$ can
be obtained by solving ${\cal L}_{\rm bFP}(x,y)[P^{{\cal S}}(x,y)\tau^{\cal S}(x,y)]=-P^{{\cal S}}(x,y)$, where $P^{{\cal S}}(x,y)$ 
is the fixation probability of the state ${\cal S}$, with the appropriate boundary conditions~\cite{Gardiner,PopGen}. Except for $\tau^{{\cal C}}$, these equations cannot be mapped onto one-dimensional problems and are difficult to solve. However, as
 $P^{{\cal AB}}\to 1$ when $w\to 1$ and $s>0$, we infer that  in this regime the equation for $\tau^{{\cal AB}}$
coincides with (\ref{MFTeq}) and therefore $\tau^{{\cal AB}}\simeq \tau$, as confirmed by Fig.~\ref{FixTime}(a). 
Similarly, since $P^{{\cal C}}\to 1$ when $w\ll 1$ and $s<0$, 
$\tau^{{\cal C}}$ coincides with $\tau$ in such a regime, see Fig.~\ref{FixTime}(b). 
The numerical results reported in Fig.~\ref{FixTime}(a,b)
show that all $\tau^{\cal S}$'s scale linearly in $N$ (and monotonically with $x+y$), in a manner that appears to be independent of
the sign of $s$. In fact, while the results for $\tau^{{\cal A}}$ and $\tau^{{\cal B}}$ in Fig.~\ref{FixTime}(b), where $s<0$,
 and those of  Fig.~\ref{FixTime}(a), where $s>0$, are comparable, the former are more noisy than the latter. 
Furthermore, as also  found for the special case $q=0$ \cite{Redner04},
the extremists MFTs, i.e. $\tau^{{\cal A}}$ and $\tau^{{\cal B}}$, are always the longest mean fixation times.
\section{Conclusion}
We have generalized a basic three-state  opinion dynamics model introduced  in Refs.~\cite{VKR03,Redner04}
and considered a finite population of $N$ individuals consisting of leftists, rightists and centrists 
interacting 
on a complete graph.
 Motivated by recent studies concerning the formation of cultural diversity, the 
system's dynamics is characterized by two competing features: (i)  ``extremists'' 
(leftists $A$'s or rightists $B$'s) 
interact with centrists ($C$'s) and at each elemental step an extremist can become a centrist with  rate $(1-q)/2$,
while a centrist can become either leftist or rightist with  rate $(1+q)/2$; (ii) 
extremists do not interact.
The former feature drives the system towards {\it consensus} while the latter accounts for 
{\it incompatibility} and leads
to a frozen steady state  of $A$'s and $B$'s (polarization).
The parameter $q$  denotes a bias favoring polarization when $q>0$ and centrism when $q<0$. 

This  three-species voter model is characterized by three absorbing fixed points (one for each species) 
and an absorbing line where  a frozen mixture of extremists coexist (polarization). 
Here, we have studied the model's final state properties
and showed that fluctuations drastically alter the mean field predictions in the presence of a small bias. 
In fact, while polarization is generally the most probable outcome when $q>0$ there still is
 a finite probability to attain a consensus, and the opposite situation arises when $q<0$.
Our results are particularly relevant in the limit of large population size
$N \gg 1$ and vanishing bias $|q|\ll 1$ with $s\equiv Nq={\cal O}(1)$ finite, where the
 fluctuations and the deterministic drift are of the same intensity.
Such a situation corresponds to the ``weak selection limit'' frequently considered
 in life and behavioral sciences~\cite{PopGen,EGT1,EGT2}.

The polarization fixation probability, $P^{{\cal AB}}$, has been
obtained analytically by solving a (separable)  backward Fokker-Planck equation and our  results have been checked 
against stochastic simulations.   Via a  mapping onto a population genetics model,  
we have shown that when $s<0$ $P^{{\cal AB}}$ approximately grows exponentially with the initial 
density $x+y$ of extremists, whereas the centrist fixation probability $P^{{\cal C}}$ decays (approximately) exponentially with $x+y$ when $s>0$.
 The mean fixation times (MFTs) have  been computed and found to scale linearly with $N$ (see Fig.~\ref{FixTime}). Furthermore, the unconditional MFT $\tau$ has been shown to 
satisfy 
$\tau(x+y,s)=\tau(1-x-y,-s)$.

Our results  show that mean field rate equations cannot describe the final state of
a simple opinion-dynamics model in the presence of a small bias.
This further illustrates the pertinence of statistical 
physics methods to describe the evolutionary  dynamics of models of cultural diversity.
\\
{\bf Acknowledgments:} Sid Redner is gratefully acknowledged for a  useful discussion.


\begin{thebibliography}{99}
\bibitem{FixProb}
The term ``fixation probability of state ${\cal S}$'' is adopted from population genetics~\cite{PopGen} and 
refers to the probability that  the system  
final state is the absorbing state ${\cal S}$.
\end{thebibliography}
\end{document}